\documentclass[pdflatex,sn-mathphys-num]{sn-jnl}% Math and Physical Sciences Numbered Reference Style

\usepackage[T1]{fontenc}
\usepackage{graphicx}%
\usepackage{multirow}%
\usepackage{amsmath,amssymb,amsfonts}%
\usepackage{amsthm}%
\usepackage{anyfontsize}
\usepackage{mathrsfs}%
\usepackage[title]{appendix}%
\usepackage{xcolor}%
\usepackage{textcomp}%
\usepackage{manyfoot}%
\usepackage{booktabs}%
\usepackage{algorithm}%
\usepackage{algorithmicx}%
\usepackage{algpseudocode}%
\usepackage{listings}%
\usepackage{comment}
\usepackage{soul}
\usepackage{subcaption}
\usepackage{ltablex}  
\usepackage{marginnote}

\theoremstyle{thmstyleone}%

\theoremstyle{thmstyletwo}%

\theoremstyle{thmstylethree}%

\begin{document}

\title[Bridging the Omics Divide: A Modular Relational Approach to Multi-Layer Biological Data Management]{Bridging the Omics Divide: A Modular Relational Approach to Multi-Layer Biological Data Management}

\author*[1]{\fnm{Alessandro} \sur{Balestrucci}}\email{alessandro.balestrucci@um.edu.mt}

\author[1]{\fnm{Donald} \sur{Friggieri}}\email{donald.friggieri@um.edu.mt}

\author[1]{\fnm{Andrea} \sur{Gariboldi}}\email{andrea.gariboldi@um.edu.mt}

\author*[1]{\fnm{Panagiotis} \sur{Alexiou}}\email{panagiotis.alexious@um.edu.mt}

\affil*[1]{\orgdiv{Centre for Molecular Medicine and Biobanking}, \orgname{University of Malta}, \orgaddress{\city{Msida}, \postcode{MSD 2080}, \country{Malta}}}

\abstract{
    \textbf{Background:} The rapid growth of high-throughput molecular data has created a critical need for data management systems that support efficient retrieval, integration, and scalability across multiple omics layers. Traditional file-based workflows, while widely used, hinder cross-modal analysis and reproducibility due to fragmented storage and ad‐hoc querying. Although several tools exist for storing and querying genomic variation data, few are designed with modular multi‑omics integration as a primary goal. To address this gap, we developed \textit{vcf2db}, a sample‑centric relational framework that models each omics modality as a distinct but linkable component, centred on biological samples. This study evaluates whether such a design can deliver competitive retrieval performance for genomic data while enabling straightforward extension to additional molecular layers.

    \textbf{Results:} We implemented a proof‑of‑concept genomic schema and ingestion pipeline for annotated VCF data, using the European subset of the 1000 Genomes Project (502 samples, 25 million variants). The system was benchmarked against three established VCF oriented tools across seven representative retrieval tasks, including coordinate‑based filtering, annotation‑driven queries, genotype extraction, and aggregation workloads. Under controlled conditions, \textit{vcf2db} demonstrated strong performance in selective queries, often outperforming the other systems for coordinate and annotation filters, and remained practically usable for genotype retrieval. Aggregation‑heavy tasks were less efficient, highlighting clear directions for future optimization. We further validated modular extensibility by incorporating a synthetic transcriptomic layer without modifying the original genomic tables, demonstrating seamless cross‑layer linking through shared sample identifiers.

    \textbf{Conclusion:} Beyond data retrieval performance, we looked at the ability to express cross-layer retrieval tasks directly as SQL queries anchored on shared sample identifiers, enabling integrated multi-omics data access that is difficult to achieve with file-based approaches.
}

\maketitle

\section{Introduction}
\label{sec:newintro}

The rapid evolution of high-throughput experimental technologies has profoundly transformed life-science research, enabling the simultaneous measurement of multiple molecular layers to capture the complexity of biological systems and disease mechanisms in a holistic manner. However, the growth in the volume, velocity, and heterogeneity of biological data, poses substantial challenges for data storage, organization, management and data/information retrieval \cite{Nardini2015Editorial}.

Such a combination produces complex digital assets that are difficult to manage using traditional file-based approaches \cite{Cuellar2022A}. While early bioinformatics workflows relied on ad-hoc scripts and collections of flat files, such solutions struggle to scale in the presence of large, evolving, and multi-layered datasets, particularly when long-term reuse, integration, and reproducibility are required \cite{Krassowski2020State}.

These limitations become especially evident where meaningful biological insights often emerge only through cross-layer integration and comparative analyses across samples and cohorts. Fragmented scripts and loosely organized files are typically error-prone, difficult to maintain, hard to query, and poorly suited for collaborative research, leading to data silos and underutilization of valuable experimental results \cite{Conesa2019Making}. Moreover, such approaches rarely provide systematic support for metadata management, standardized annotations, access control, or compliance with FAIR data principles, all of which are increasingly expected in contemporary biomedical research \cite{Doniparthi2024Integrating}.

In response to these challenges, several studies have emphasized the necessity of centralized, database-driven solutions for multi-omics data management. Centralized platforms enable structured data modeling, consistent metadata representation, efficient indexing, and flexible querying across heterogeneous omics layers, which are not feasible with purely file-based approaches \cite{Neogi2013MoDa-A}. By providing a unified view of multi-omics datasets, such systems facilitate data integration, reproducibility, and collaborative analysis, while supporting scalable data ingestion and long-term data stewardship \cite{Cuellar2022A,Krassowski2020State}.

Within this broader context, genomic variation data provides a particularly demanding and informative test case representing one of the most challenging omics data types in terms of scale, structural complexity, and sample-level dimensionality. For this reason, we think that solutions tailored to store and retrieve VCF data represent a crucial starting point to evaluate if a new design can support not only efficient retrieval, but also progressive multi-omics integration.

At the best of our knowledge, existing approaches for VCF management can be broadly grouped into two categories. The first includes database-backed solutions that ingest VCF data into relational or non-relational systems. Among the tools considered in this work, VCFdbR \cite{koomar2020vcfdbr} stores information in SQLite and offers alternative strategies for genotype handling, while Cutevariant \cite{schutz2022cutevariant} combines a local SQLite backend with a graphical interface and a domain-specific query language for annotated VCF exploration. 
VCF-Miner instead adopts MongoDB to support filtering and interactive exploration of variant and annotation data \cite{hart2016vcf}. Although these tools demonstrate VCF data can be queried effectively with different storage engines and schema designs, they are primarily oriented toward variant exploration and mining, rather than multi-omics data integration.
The second category includes file-based approaches, which avoid database ingestion and operate directly on VCF or VCF-derived representations. VCFtools established the foundations for parsing and analysing VCF files, and subsequent tools such as VIVA \cite{tollefson2019viva}, re-Searcher \cite{karabayev2021re}, BrowseVCF \cite{DBLP:journals/bib/SalatinoR17}, and VCF Zarr \cite{czech2025analysis} further improved direct access, filtering efficiency, or storage layout for large-scale datasets. These approaches are valuable because they eliminate schema-definition overhead and can be highly effective for specific access patterns. At the same time, beside the mentioned limitation of using a file-based approach\cite{Conesa2019Making} they generally leave cross-layer integration dependent on external workflows, repeated preprocessing, or modality-specific scripting.

The main gap, therefore, is not the absence of tools for variant storage or retrieval, but the lack of a data-management methodology that combines competitive genomic querying with an explicit design for modular multi-omics integration. This is the problem addressed in this study. 
Rather than proposing only another VCF querying tool, our goal is to define a relational and sample-centric framework in which each omics modality can retain its own internal organization while remaining directly linkable through a shared biological backbone. In such a model, \texttt{sample} entities become the stable integration point through which genomics, transcriptomics, and future molecular layers can be connected using explicit relational constraints instead of external file matching or ad-hoc identifiers. To properly define and evaluate the usefulness and effectiveness of our proposal, the following research questions guided our work:

\begin{itemize}
\item Can annotated VCF data be decomposed into normalized relational entities in a way that preserves biologically meaningful querying while maintaining efficient retrieval for common genomic tasks?
\item Can a sample-centric relational schema support direct and modular extension beyond genomics, so that additional omics layers can be incorporated without redesigning the original genomic component?
\item Can such a model remain competitive with existing VCF-oriented systems across representative retrieval workloads while at the same time providing a stronger basis for cross-layer querying?
\end{itemize}

In this work we propose \textit{vcf2db}, a relational database based framework able to models each omics modality as a distinct but linkable component within a star-like architecture centred on biological samples. In the genomic implementation, annotated VCF information is decomposed into tables representing genomic loci, alternative alleles, functional annotations, genotype observations, and sample-level relationships. The advantage of applying a normalization strategy is twofold: (i) to reduce redundancy and (ii) to align the data structure with the way genomic questions are actually asked (e.g., coordinate-based searches target loci, functional filtering targets annotations, and cohort-aware analyses target sample-linked genotypes). To support practical retrieval performance, the schema is combined with chromosome-based partitioning of the main variant table and indexes tailored to genomic coordinates and frequently queried annotation fields. 
To evaluate the scope and practical performance of the proposed framework, we implemented a proof-of-concept Python ingestion pipeline\footnote{\url{https://github.com/BioGeMT/vcf2db}}. The pipeline parses annotated VCF records iteratively, maps them to the relational schema, and inserts them in batches. We tested the implementation using annotated genomic variation data from the European subset of the 1000 Genomes Project Phase 3 release, comprising 502 samples and 25,057,597 variants. Additional 1\%, 10\%, and 50\% sample subsets were also generated to assess scaling behaviour.
Benchmarking was designed around seven representative genomic retrieval tasks covering lightweight coordinate- and annotation-driven filtering, genotype-level retrieval, and aggregation-heavy analytical workloads. The proposed implementation was compared against three established VCF-oriented solutions, namely Cutevariant, VCFdbR, and VCF-Miner. These tools were selected because they support storage and querying of positions, alleles, genotypes, and annotations while relying on different backend and schema strategies. All experiments were executed on the same high-performance computing server under controlled conditions, using repeated runs and focusing primarily on cold-cache execution after clearing the operating system page cache and restarting services, with warm-cache runs also recorded to assess cache sensitivity. This design allowed us to compare not only absolute query times on the full dataset, but also workload-specific behaviour and runtime scaling as the number of samples increased.

Furthermore, since the aim was not only to store genomic variation but to demonstrate modular expansion, the framework was also extended with a transcriptomic layer. Synthetic RNA sequencing-like gene expression data were generated, ingested and linked to the same sample entities used for genomics, showing that a new molecular modality could be added without modifying the existing genomic tables. This design choice is central to the contribution of the work: modality-specific data remain independently structured and optimizable, while cross-modality integration is achieved through a stable sample backbone. The main contributions of this study can be summarized as follows:
\begin{itemize}
\item a sample-centric methodology (\textit{vcf2db}) for omics data organization and retrieval, treating multi-omics integration as a database-design problem rather than as a downstream scripting task;
\item a normalized genomic schema and ingestion strategy to support efficient retrieval;
\item a proof of concept of the modular extensibility of the framework by incorporating a transcriptomic layer;
\item a benchmark to compare the performance of our system against other solutions across representative genomic retrieval workloads. 
\end{itemize}

The experimental results show that \textit{vcf2db} performed particularly well in selective retrieval workloads, including annotation-driven filtering and coordinate-based queries, and remained practically usable for genotype-level extraction. 
The benchmark also revealed specific efficiency bottlenecks in the current implementation, most notably in full-dataset aggregation and multi-level summarization, presenting key challenges for future optimization.
These findings are important because they show that the framework is not only structurally robust and extensible but also operationally competitive on large annotated genomic datasets; demonstrating that a carefully designed relational model can serve as a powerful, scalable bridge between fragmented data silos and the integrated, queryable knowledge needed to unlock new biological insights. 

The remainder of the paper is organized as follows. Section \ref{sec:methodsAB} describes the sample-centric schema, the genomic and transcriptomic modelling choices, the ingestion pipeline, and the benchmark design. Section \ref{sec:resultsAB} reports retrieval performance, cache effects, and runtime scaling across dataset fractions. Section \ref{sec:disc} interprets these findings in relation to alternative systems, highlights the strengths and limitations of the current implementation, provides an answer to our research questions, and outlines directions for optimization and real multi-omics deployment. Section \ref{sec:conc} summarizes the methodological contribution of vcf2db as a modular and extensible framework for scalable omics data management.

\section{Methods}
\label{sec:methodsAB}

To address the research questions outlined in the introduction, this section presents the proposed relational approach for omics data management, using VCF genomic data as the initial use case and proof of concept.

We describe the rationale behind the main design choices, the resulting sample-centric database architecture, and its extension to an additional omics layer through the generation and integration of transcriptomic data. 
Finally we describe the experimental setup, including the datasets, alternative solutions, and benchmarking strategy used to evaluate retrieval performance.

\subsection{The sample-centric design}
\label{subsec:schemadesign}

The proposed database model was designed around biological samples\footnote{A 'sample' is defined as any biological entity (in this case a human individual or a patient) for which one or more omics measurements are available.}
as the primary unit of organization and integration. This choice reflects a central requirement of multi-omics studies: different molecular measurements are often generated from the same individual, specimen, or experimental condition but are stored in heterogeneous formats and produced at different stages of the analytical workflow. By anchoring all data modalities to a shared samples entity, the schema provides a stable integration backbone that enables genomic, transcriptomic, phenotypic, and other omics data to be linked through explicit relational constraints rather than through external file matching or ad hoc identifiers.

\begin{figure}[htbp]
    \centering
    \includegraphics[width=\linewidth]{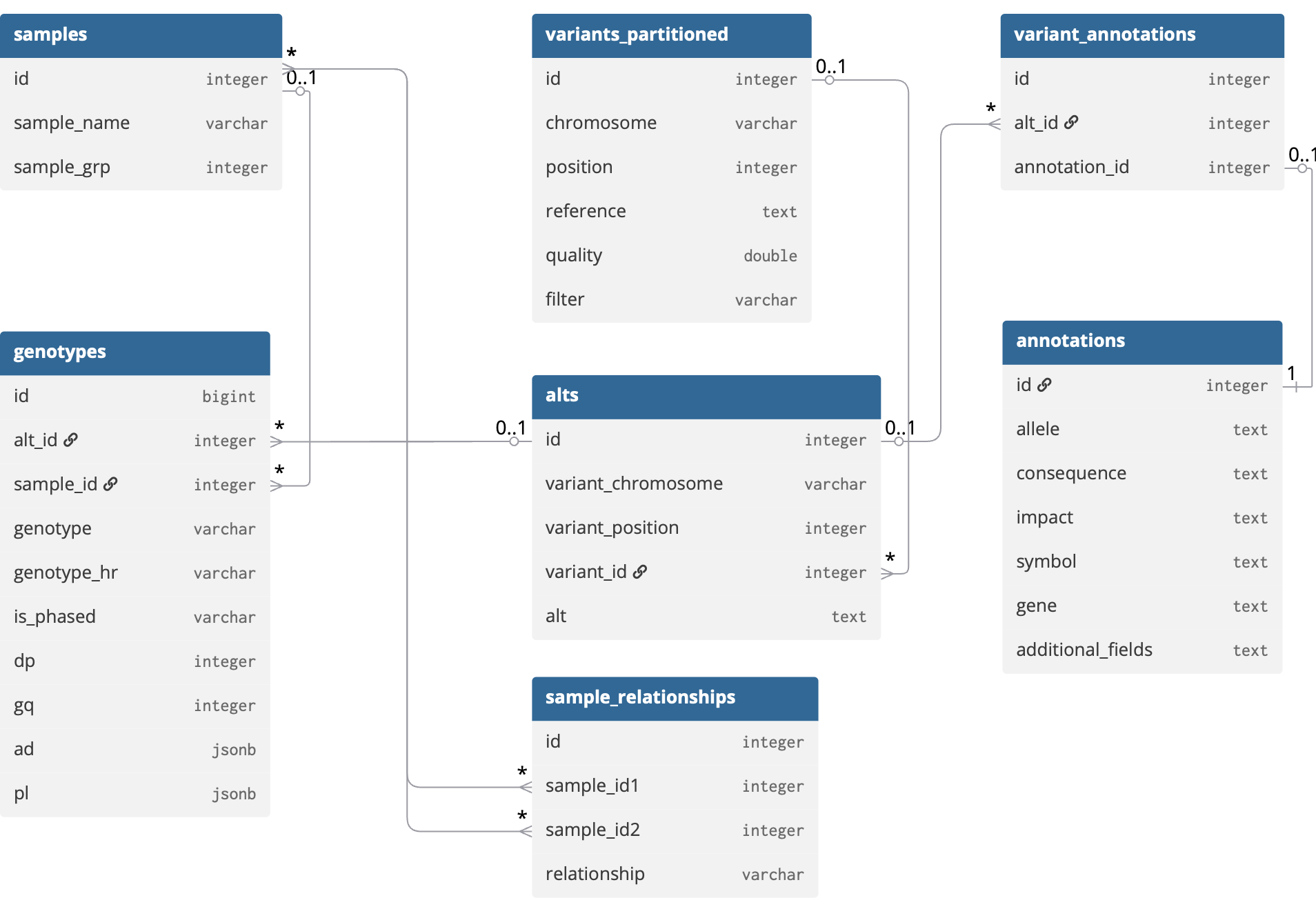}
    \caption{Entity-relationship diagram illustrating the sample-centric database schema, where the \textit{samples} table serves as the central entity.}    \label{fig:erd1}
\end{figure}

Figure \ref{fig:erd1} shows the Entity–Relationship model where the \textit{samples} table is linked to other tables related to genomic data.

The samples table stores a unique identifier for each biological sample, which serves as a foreign key for linking sample-level metadata (such as cohort or group membership) stored in related tables.
The use of a sample entity allows all downstream molecular observations to be linked to the same biological unit, thereby supporting comparative and integrative analyses across samples and data modalities. Additional tables, such as \textit{sample\_relationships}, provide support for cohort stratification and for representing known relationships between samples, including family relationships, technical replicates, or other study-specific dependencies. 

For the genomic data modality, the information was decomposed into multiple relational entities in order to represent distinct biological concepts at the appropriate level of granularity. Genomic loci are stored in the \textit{variants\_partitioned} table, which records chromosome, genomic position, reference allele, quality score, and filter status. Alternative alleles are stored separately in the \textit{alts} table and linked to their corresponding variant locus through foreign keys. This separation is important because a single genomic position may contain more than one alternative allele; modeling alleles as independent records avoids ambiguity and provides explicit support for multi-allelic variants.

Functional annotations are represented independently from both variant loci and genotypes and stored in the dedicated \textit{annotations} table. Since a single allele may have multiple functional annotations, for example across different transcripts or regulatory features, the schema uses an associative table, \textit{variant\_annotations}, to link alternative alleles to one or more annotation records. This design avoids storing the same annotation repeatedly for every sample and allows queries to target biological attributes such as gene symbol, consequence type, predicted impact, or transcript-level effects.

Sample-specific genotype information is stored in the \textit{genotypes} table, which represents the relationship between samples and alternative alleles. This table captures the occurrence and characteristics of each allele in each sample, including genotype call, phasing status, depth, genotype quality, allelic depth, and likelihood-related
information when available. Conceptually, this table is the bridge between the sample-centric part of the schema and the variant-centric part of the schema: variants and alleles describe what genomic events exist, whereas genotypes describe how those events are distributed across individuals. 

A key design principle of the schema is normalization. Indeed, rather than storing variants information as a single wide record, the model separates these components into dedicated tables. This deceases redundancy, improves consistency, and enables different query patterns to operate on the relevant level of biological information. For example, coordinate-based searches can be performed on variant loci, consequence-based searches on annotations, and cohort-based genotype analyses on the genotype table. These tables remain connected through primary and foreign keys, allowing complex retrieval tasks to combine genomic position, allele identity, functional consequence, sample group, and genotype attributes within a single query.

To improve scalability, the main variant table was implemented using a chromosome-based partitioning strategy. The \textit{variants\_partitioned} table is divided into separate physical partitions according to chromosome, including autosomes, sex chromosomes, mitochondrial DNA, and a default partition for non-standard chromosome identifiers. This choice reflects a common access pattern in genomic analyses, where queries frequently target specific chromosomes or genomic intervals. When a query includes a chromosome or chromosomal region, the database management system engine can apply partition pruning and restrict execution to the relevant partition instead of scanning the entire variant table. This reduces the amount of data accessed during query execution and improves performance for region-specific retrieval tasks.
Partitioning also improves index management, as each chromosome partition maintains smaller local indexes, for example on genomic position. Compared to a single monolithic table, these indexes are easier to cache and faster to scan during coordinate-based queries. Chromosome-level partitioning was chosen as a biologically meaningful compromise between performance and schema complexity as it enables partition pruning for region-specific queries while avoiding the overhead of managing many small partitions. 
Queries without chromosome filters may still scan multiple partitions, but most genomic retrieval tasks are naturally defined by chromosomes, genes, or genomic intervals.
To further support efficient data retrieval, the schema was complemented by an indexing strategy tailored to the main query patterns expected in genomic and multi-omics analyses. Primary and foreign key indexes support joins across samples, genotypes, alleles, variants, and annotations, while dedicated indexes on genomic coordinates enable efficient region-based searches. Additional indexes were defined on frequently queried annotation fields, such as gene identifiers and consequence terms, to accelerate functional filtering. For partitioned variant tables, indexes are maintained locally within each chromosome partition, allowing coordinate-based queries to benefit from both partition pruning and index scans. This indexing strategy was designed to balance retrieval performance with storage overhead, ensuring that the database remains suitable for large-scale exploratory and integrative analyses.

\subsection{Extensibility for multi-omics integration}

The central objective of the proposed schema is to support the progressive integration of heterogeneous omics layers without requiring a redesign of the original database structure. The schema is therefore organized around a stable sample-centric backbone, in which the \textit{samples} table acts as the common reference entity for all molecular and phenotypic measurements. This design reflects the structure of multi-omics studies, where genomics, transcriptomics, metabolomics, proteomics, clinical phenotypes, and other data types are often generated from the same biological samples but differ substantially in format, dimensionality, and analytical meaning.

\begin{figure}[htbp]
    \centering
    \includegraphics[width=\linewidth]{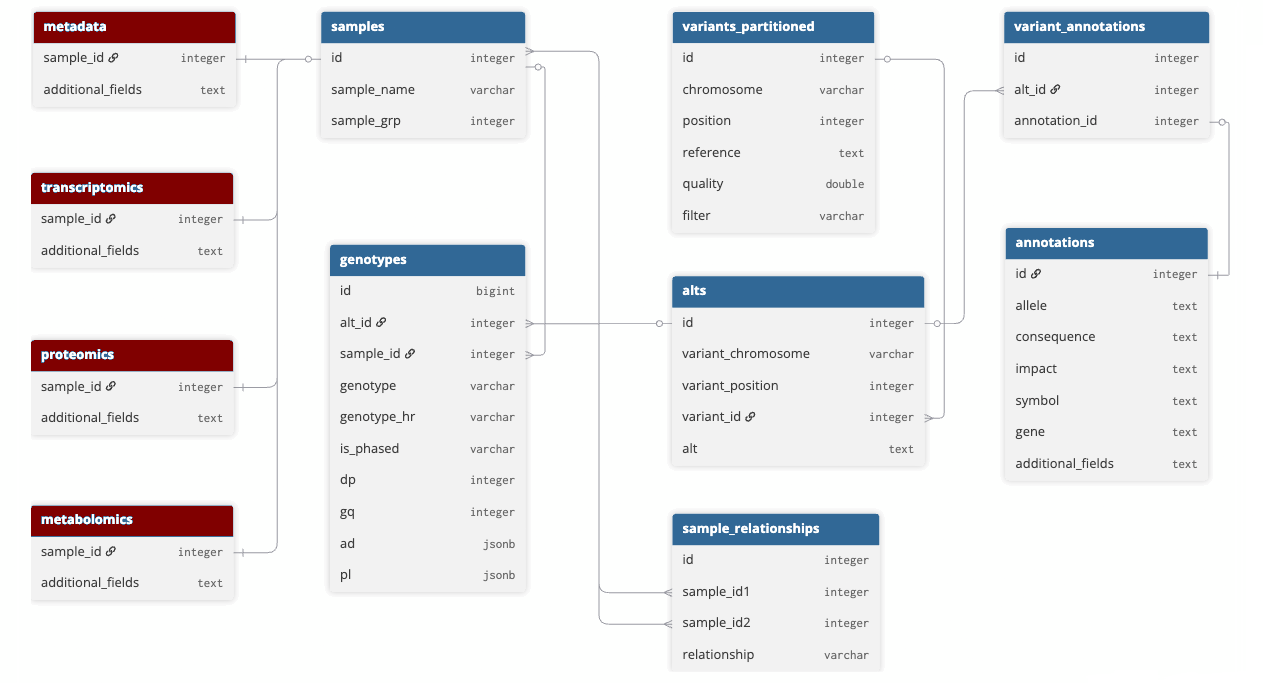}
    \caption{Conceptual extension of the core database schema with supplementary tables
             (shown in red) representing additional omics layers and sample-level metadata.}
    \label{fig:schema_ext2}
\end{figure}
As illustrated in Figure \ref{fig:schema_ext2}, extensibility can be easily achieved by adding modality-specific tables that reference the central samples table through foreign keys. Each new omics layer can therefore be modeled according to its own biological structure while remaining directly connected to the rest of the database. 
For example, Figure \ref{fig:schema_ext2} shows transcriptomics measurements can be stored in dedicated expression tables, metabolomic profiles in metabolite-level tables, and clinical or other important data in sample-level \textit{metadata} tables. This organization avoids forcing heterogeneous data types into a single generic structure and preserves the specific characteristics of each modality.

In this work, the extensibility of the schema is demonstrated by extending the initial based genomic model with transcriptomics data. Gene-level expression measurements were represented in additional tables linked to the same samples used for genomic variation data. This extension did not require modifications to the existing variant; instead, the transcriptomics layer was incorporated as an independent module connected through the shared sample identifier. This shows that the schema can evolve incrementally as new data modalities become available.

The main advantage of this approach is that multi-omics retrieval can be performed directly at the database level. 
For instance, researchers can retrieve samples carrying variants in a specific gene and, in the same query, inspect the expression level of that gene across the corresponding individuals. 

Overall, the extensible schema separates two complementary requirements of multi-omics data management: modality-specific representation and cross-modality integration. Each omics layer remains independently modeled, indexed, and maintained, while the sample-centric backbone provides a consistent mechanism for linking data across layers. This makes the database suitable not only for the genomic use case presented here, but also as a scalable foundation for future extensions to additional omics technologies and sample-level metadata.

\subsection{Required Materials - Proof of concept}
\label{subsec: proof_concept}
The proof of concept developed in this work was built using genomic variant data as the initial test case for the proposed relational schema. Specifically, we used data from the 1000 Genomes Project Phase 3 release, which originally contains approximately 78 million variants across 2,504 samples. To obtain a dataset that was computationally tractable while still representative of large-scale genomic studies, we restricted the analysis to individuals belonging to the European (EUR) super-population, resulting in 502 samples. Since the original release does not include functional annotations, the dataset was annotated using the Ensembl Variant Effect Predictor (VEP), version 113, with the \textit{-\--everything} option, thereby enriching the VCF records with information from Ensembl, RefSeq, regulatory features, and population frequency resources. The final annotated dataset contained 25,057,597 variant records across 502 samples, stored as a compressed VCF file of approximately 77 GB.

To evaluate scalability, three additional datasets were generated by randomly subsampling the 502 European samples while preserving the complete set of variants. These subsets corresponded to 1\%, 10\%, and 50\% of the cohort, resulting in datasets containing 5, 50, and 251 samples, respectively. Together with the full 502-sample dataset, these subsets allowed us to assess how retrieval performance changed as the number of samples increased, while keeping the variant space constant. This design was particularly useful for evaluating genotype-related queries, whose computational cost is expected to scale with the number of sample-specific genotype records.

The annotated VCF files were ingested into the proposed database using a dedicated Python-based ingestion pipeline\footnote{\url{https://github.com/BioGeMT/vcf2db}}. The pipeline parses VCF records through an iterator-based strategy, avoiding the need to load the full input file into memory. During ingestion, each VCF record is decomposed according to the relational schema: genomic loci are inserted into the partitioned variant table, alternative alleles are stored in a dedicated allele table, functional annotations extracted from the VEP CSQ field are stored separately, and sample-specific genotypes are linked to the corresponding samples and alleles. To improve scalability, records are processed in configurable batches and inserted using batch operations, reducing database roundtrips and controlling transaction size. The ingestion procedure also supports parallel batch processing, with independent database connections for workers, and includes error logging for malformed annotations, missing fields, or inconsistent genotype information. This allowed the import process to remain robust even in the presence of imperfect real-world VCF records.

To demonstrate the extensibility of the schema beyond genomics, we generated and integrated a transcriptomic layer associated with the same sample entities. Synthetic RNA-seq-like data were generated at gene level and stored in additional tables linked to the central samples table. The transcriptomic data included both raw read counts and normalized expression values for each gene–sample pair. Gene expression values were generated using biologically inspired expression profiles, including highly expressed, medium expressed, low expressed, very low expressed, and non-expressed genes. Detection rates and stochastic variability were introduced to approximate realistic RNA-seq distributions. The generation procedure also allowed differential expression patterns to be simulated between sample groups by selecting a subset of genes and applying fold changes. This transcriptomic extension was incorporated without modifying the existing genomic tables, thereby providing a practical demonstration of the modularity of the sample-centric schema.

Benchmarking was designed to evaluate retrieval performance under realistic genomic data access patterns. The proposed implementation was compared against three existing VCF-oriented database solutions: Cutevariant \cite{schutz2022cutevariant}, VCFdbR \cite{koomar2020vcfdbr}, and VCF-Miner \cite{hart2016vcf}. 
These tools were selected because they support the storage and querying of VCF-compliant information, including genomic positions, alleles, genotypes, and annotations, while representing different database backends and schema design strategies. All experiments were executed on the same high-performance computing server, equipped with 48 CPU cores and 755 GB of memory, and all database systems were configured to use the same local storage in order to minimize differences due to hardware or I/O configuration.

The benchmark included seven representative retrieval tasks, selected to cover common operations in genomic analysis pipelines and listed in Table \ref{tab:query-descriptions1}

\begin{table*}[htbp]
\centering
\footnotesize
\setlength{\tabcolsep}{4pt}
\caption{Description of the benchmark queries along with their number of rows returned.}
\label{tab:query-descriptions1}
\begin{tabular}{clr}
\toprule
Query & Description & Rows \\
\midrule
Q1 & Variants in genomic coordinates (chr1:100000--200000) & 51 \\
Q2 & Variants filtered by gene (BRCA1) and consequence (\textit{missense\_variant}) & 24 \\
Q3 & Rare variants (AF $<$ 0.01) with high functional impact on chromosomes 1--3 & 3{,}950 \\
Q4 & Per-sample genotypes in genomic coordinates (chr17:41000000--42000000) & 4{,}063{,}690 \\
Q5 & Variants with quality $>$ 20 and multiple alternate alleles, grouped by position & 86{,}491 \\
Q6 & Top 50 protein-coding genes with the most variants, aggregated by gene and chromosome & 50 \\
Q7 & Chromosome-wide statistics: variant count, annotation coverage, and average quality & 25 \\
\bottomrule
\end{tabular}
\end{table*}

For each database solution and dataset size, the benchmark queries were executed under controlled conditions. Query execution was evaluated primarily under cold-cache conditions, which represent a demanding and realistic scenario in which data are not already loaded into memory. Before cold-cache runs, the operating system page cache was cleared and database services were restarted to remove residual cached data. Each query was executed three times, and the mean execution time was reported to reduce the influence of transient system variability. Warm-cache executions were also considered, but the relative ranking between systems did not substantially differ from the cold-cache results.
This benchmarking procedure allowed us to compare both absolute query performance and scaling behaviour as the number of samples increased.
\section{Results}
\label{sec:resultsAB}

Building on the benchmarking strategy described in Section \ref{subsec: proof_concept}, we report the performance of our sample-centric relational schema across seven representative query tasks (Table \ref{tab:query-descriptions1}). To enable a structured comparison, queries were grouped into three classes reflecting increasing computational complexity: (i) lightweight retrieval and annotation-driven filtering (Q1–Q3), (ii) genotype retrieval and variant-level analytical queries (Q4–Q5), and (iii) aggregation-heavy workloads (Q6–Q7). The analysis compares execution times across multiple database backends, systematically examining three dimensions of practical relevance: full-dataset performance under both cold- and warm-cache conditions, the differential impact of cache state on query latency, and runtime scaling as a function of dataset fraction. 
Results are presented by query class, highlighting how coordinate-driven filtering, allele-level operations, and population-scale aggregation impose distinct computational demands, thereby revealing workload-specific trade-offs rather than a single ordering of systems.

\subsection{Full-dataset benchmark performance across query classes}
\label{subsc:FullDBench_Q}

\begin{figure}
    \centering
    \includegraphics[width=1\linewidth]{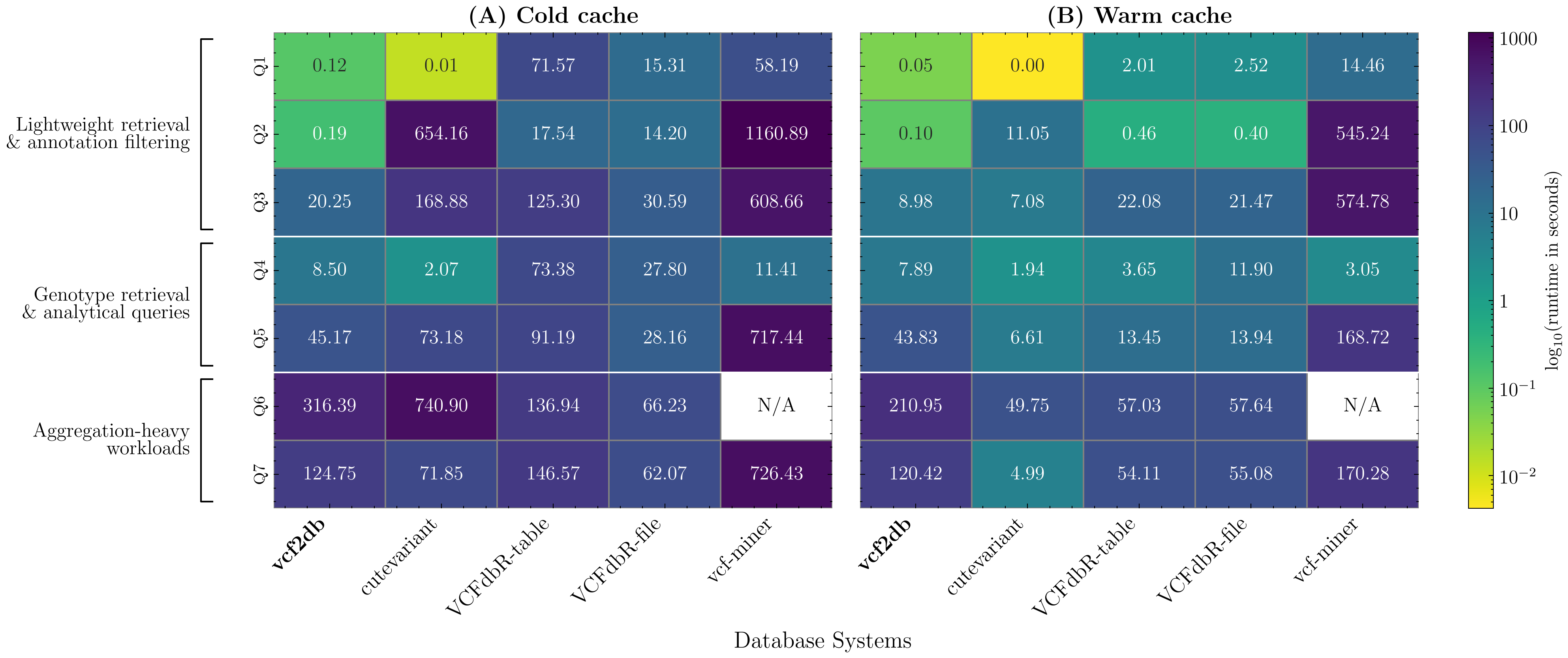}
    \caption{Full dataset query runtimes for the five variant database systems. Execution times are reported separately by (A) Cold cache and (B) Warm cache.}
    \label{fig:placeholder1}
\end{figure}
A global overview of execution times across all systems and queries on the full dataset is presented in Figure \ref{fig:placeholder1}, which summarizes mean runtimes under cold- and warm-cache conditions on a logarithmic scale. Performance differences span several orders of magnitude and are not uniform across query classes; each system exhibits a distinct profile shaped by the type of operation performed. 
Across all queries, \textit{vcf2db} showed its strongest performance in the lightweight retrieval class, where runtimes were competitive or superior to alternative systems. In contrast, genotype-level and aggregation workloads led to substantially increased execution times and a diminished relative advantage, indicating that performance is primarily determined by the interaction between query structure and the underlying data access and processing strategies.

\subsubsection{Lightweight retrieval and annotation-driven filtering}
Queries Q1–Q3 represent highly selective workloads combining coordinate-based access and annotation-level filtering. Across this class, vcf2db consistently achieved the lowest or near-lowest execution times, particularly under cold-cache conditions.

In Q1, all systems benefit from the limited output size (recall to Table \ref{tab:query-descriptions1}), but differences in access strategy remain evident. \textit{vcf2db} completed the query in 0.116 s (cold) and 0.050 s (warm), closely matching the fastest solution (i.e., \textit{cutevariant}: 0.013 s cold; 0.004 s warm), while outperforming \textit{VCFdbR} and \textit{vcf-miner} by one to three orders of magnitude in cold-cache execution. This indicates that indexed coordinate-based access is efficiently supported across multiple systems, with \textit{vcf2db} remaining competitive with acceptable time despite not always achieving the absolute minimum latency.

The advantage of \textit{vcf2db} becomes more pronounced in annotation-driven queries. In Q2, \textit{vcf2db} was the fastest solution under both cache conditions, with mean runtimes of 0.192 s (cold) and 0.097 s (warm). In contrast, alternative systems exhibited substantially higher cold-cache costs, reaching 17–654s depending on the implementation, and exceeding 1000s for \textit{vcf-miner}. This large gap suggests that join-heavy annotation filtering is handled particularly efficiently by our solution.

A similar pattern is observed in Q3, where vcf2db achieved the lowest cold-cache runtime (20.246 s), outperforming \textit{VCFdbR} (30–125 s) and \textit{vcf-miner} (608 s). Under warm-cache conditions, \textit{vcf2db} (8.976 s) remained competitive with the fastest system (\textit{cutevariant}: 7.082 s).

Overall, these results indicate that \textit{vcf2db} is particularly well suited for selective queries combining genomic coordinates and functional annotations, where indexing and normalized schema design enable efficient filtering and join execution.

\subsubsection{Genotype retrieval and variant-level analytical queries}

Queries Q4 and Q5 introduce substantially larger intermediate result sets and require operations at the genotype or variant-group level, resulting in markedly higher computational costs than the lightweight retrieval class.
Q4 (per-sample genotype retrieval within a genomic interval) returns over four million rows. \textit{vcf2db} required 8.502 s (cold) and 7.889 s (warm). While remaining within a moderate execution range, it was slower than \textit{cutevariant} (2.071 s cold, 1.944 s warm) and also slower than \textit{vcf‑miner} under warm-cache conditions (3.051 s), though it significantly outperformed \textit{VCFdbR-table} in the cold-cache scenario (73.378 s). The small difference between cold and warm runtimes for \textit{vcf2db} (warm/cold ratio $\approx$ 0.93) suggests that performance is dominated by query execution and data processing rather than disk I/O.
In Q5 (multi-allelic variant identification with quality filtering and grouping), \textit{vcf2db} showed intermediate performance, with runtimes of 45.171 s (cold) and 43.828 s (warm). It was slower than \textit{VCFdbR-file} in cold-cache conditions (28.160 s) and substantially slower than \textit{cutevariant} after cache warming (6.609 s). As in Q4, the limited reduction between cold and warm execution times (warm/cold ratio $\approx$ 0.97) indicates that the computational cost of grouping and aggregation operations dominates execution.

\subsubsection{Aggregation-heavy workloads}
Queries Q6 and Q7 represent global aggregation tasks requiring full-dataset scans and summarization, and therefore constitute the most computationally demanding workloads in the benchmark.
In Q6, \textit{vcf2db} required 291.337 s (cold) and 210.955 s (warm), compared to substantially lower runtimes for \textit{VCFdbR-file} (66.226 s cold) and \textit{VCFdbR-table} (136.938 s cold), and a markedly lower warm-cache runtime for \textit{cutevariant} (49.753 s). Although cache warming reduces execution time for \textit{vcf2db} (warm/cold ratio $\approx$ 0.72), the remaining runtime indicates that aggregation and grouping over large datasets remain computationally intensive.
In Q7, \textit{vcf2db} exhibited similarly high runtimes (124.751 s cold; 120.424 s warm), with minimal improvement from cache warming (ratio $\approx$ 0.97). 
Other solutions achieved lower execution times, particularly under warm-cache conditions (e.g. 4.990 s for cutevariant), indicating more efficient handling of global aggregation tasks.

Overall, aggregation-heavy queries expose the main performance limitations of \textit{vcf2db} in its current implementation. While the system remains fully functional and produces correct results, its execution time increases substantially when queries require %full-table scans, 
large intermediate joins, or multi-level aggregation.

\subsection{Effect of cache state on query execution time}

\begin{figure}
    \centering
    \includegraphics[width=1\linewidth]{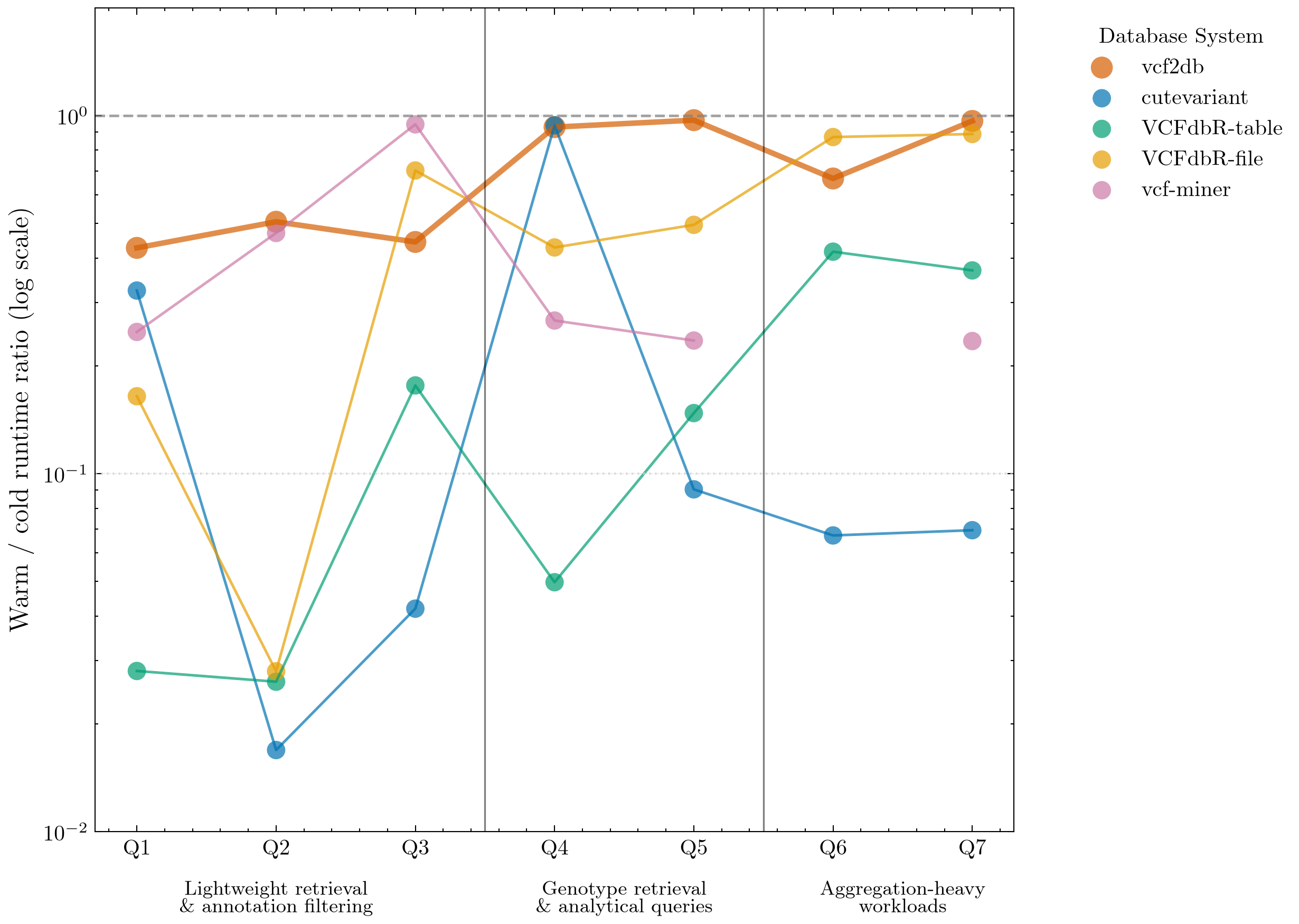}
    \caption{Warm-to-cold runtime ratios per query for the five benchmarked database systems. }
    \label{fig:placeholder2}
\end{figure}

To isolate the contribution of disk I/O to query latency, warm-to-cold runtime ratios were calculated for each system and query (Figure~\ref{fig:placeholder2}). A ratio close to 1 indicates that performance is dominated by computation and is largely insensitive to disk caching, whereas lower values reflect I/O-bound workloads that benefit substantially from memory-resident data. The effect of cache warming was markedly heterogeneous across both queries and systems, revealing fundamental differences in how each engine's execution cost is structured.

For \textit{vcf2db}, cache sensitivity was strongly partitioned by query class. Lightweight retrieval and annotation-driven queries (Q1--Q3) yielded ratios of 0.43 (Q1), 0.51 (Q2), and 0.44 (Q3), corresponding to a roughly two-fold reduction in runtime under warm-cache conditions. This demonstrates that these selective, index-driven operations are predominantly I/O-bound, with reduced disk latency directly lowering response times. In contrast, genotype-level queries (Q4--Q5) and aggregation-heavy workloads (Q6--Q7) exhibited substantially less improvement: ratios for Q4, Q5, and Q7 were 0.93, 0.97, and 0.97, respectively, indicating near-complete insensitivity to cache state. Q6 showed a moderate reduction (ratio 0.67) yet remained among the longest-running queries in absolute terms. These patterns suggest that as query complexity increases---through large intermediate result sets, genotype-level joins, and multi-level aggregation---computation rather than I/O becomes the dominant cost factor for \textit{vcf2db}.

Other systems displayed contrasting cache-sensitivity profiles. \textit{cutevariant}, for instance, achieved ratios as low as 0.02--0.09 on Q2, Q3, Q5, Q6, and Q7, pointing to a heavy reliance on disk access that was almost entirely eliminated by caching; the exception was Q4 (ratio 0.94), where genotype retrieval drove compute-bound behavior. \textit{VCFdbR-table} and \textit{VCFdbR-file} showed intermediate and query-dependent sensitivities, while \textit{vcf-miner} exhibited a mixed profile (ratios ranging from 0.23 on Q7 to 0.94 on Q3). Overall, these results indicate that cache warming disproportionately benefits lightweight retrieval workloads for \textit{vcf2db}, whereas analytical queries remain constrained by computational overhead even under ideal memory conditions.

\subsection{Runtime scaling across dataset fractions}

\begin{figure}
    \centering
    \includegraphics[width=1\linewidth]{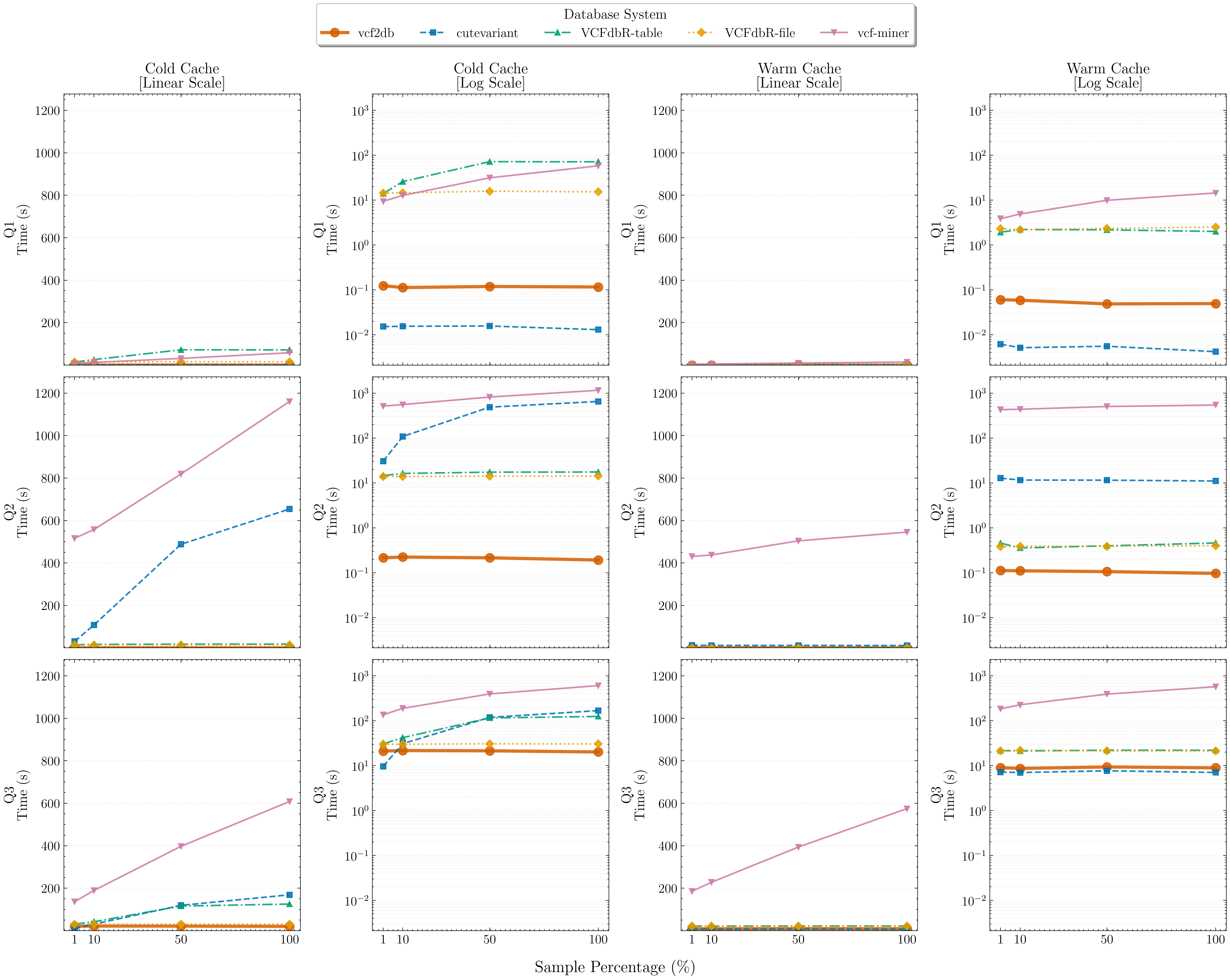}
    \caption{Query runtime scaling with the number of samples for lightweight retrieval and annotation-filtering queries (Q1-Q3).}
    \label{fig:placeholder3}
\end{figure}

\begin{figure}
    \centering
    \includegraphics[width=1\linewidth]{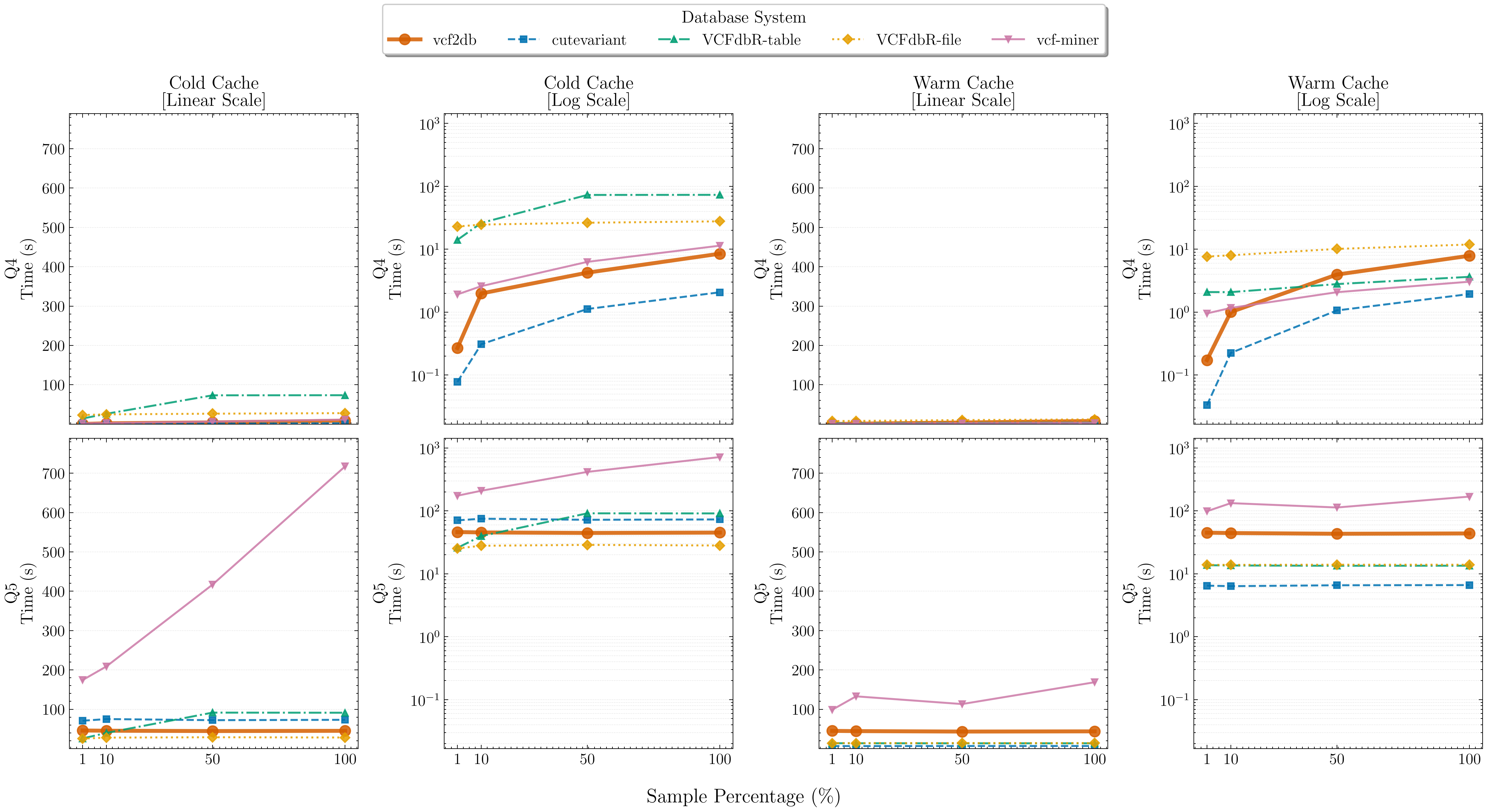}
    \caption{Query runtime scaling with the number of samples for lightweight retrieval and annotation-filtering queries (Q4-Q5).}
    \label{fig:placeholder4}
\end{figure}

\begin{figure}
    \centering
    \includegraphics[width=1\linewidth]{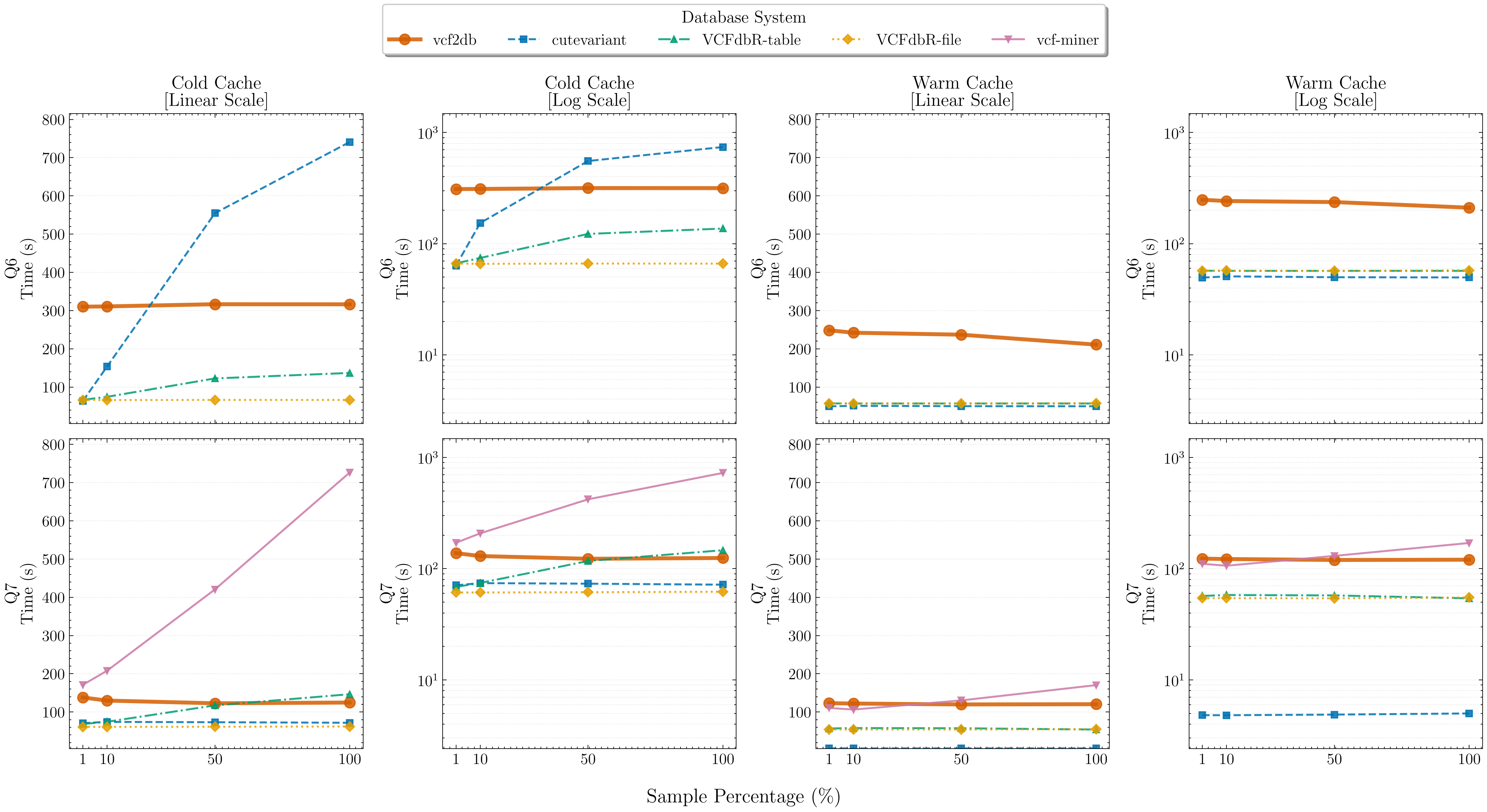}
    \caption{Query runtime scaling with the number of samples for lightweight retrieval and annotation-filtering queries (Q6-Q7).}
    \label{fig:placeholder5}
\end{figure}

To characterize how query latency responds to increasing cohort sizes, we measured execution times on subsets containing 1\%, 10\%, 50\%, and 100\% of the 502-sample 1000 Genomes EUR dataset. Results are presented separately for each query class: (i) lightweight retrieval and annotation-driven filtering in Figures~\ref{fig:placeholder3} (Q1--Q3), (ii) genotype-level analytical queries in Figure~\ref{fig:placeholder4} (Q4--Q5), and (iii) aggregation-heavy workloads in Figure~\ref{fig:placeholder5} (Q6--Q7).

Each figure includes both cold and warm cache cases and both logarithmic and linear y-axis representations: the logarithmic scale highlights multiplicative trends and relative scaling slopes, whereas the linear scale emphasizes the practical impact of runtime differences in absolute seconds and, therefore, the practical cost of larger cohorts.

\paragraph{Lightweight retrieval and annotation-driven filtering (Figure~\ref{fig:placeholder3}).}

For lightweight retrieval and annotation-driven filtering workloads (Q1--Q3; Figure~\ref{fig:placeholder3}), \textit{vcf2db} showed nearly flat scaling across dataset fractions. 

In Q1, runtimes remained consistently below 0.2~s under cold cache and close to 0.05--0.06~s under warm cache, indicating minimal sensitivity to the number of samples included in the dataset fraction. A similar pattern was observed for Q2, where \textit{vcf2db} remained close to 0.2~s under cold cache and approximately 0.1~s under warm cache across all fractions. The logarithmic representation makes this behaviour visible as almost horizontal trajectories, while the linear representation shows that the absolute runtime cost remained negligible relative to most other solutions. In contrast, \textit{cutevariant} showed strong sample-dependent growth for Q2 under cold cache, increasing from approximately 30~s at 1\% to more than 650~s at 100\%, whereas its warm-cache runtimes remained close to 11--13~s. \textit{vcf-miner} also showed pronounced growth for Q2, increasing from approximately 516~s to more than 1160~s under cold cache. 

Q3 (rare, high‑impact variants across three chromosomes) added a moderate variant‑range expansion, yet \textit{vcf2db} again showed no appreciable scaling: cold runtimes hovered around 20.2--21.8~s from 1\% to 100\%. On the log scale, the curve is horizontal, while the linear scale reveals that the absolute cost stays within a narrow band of approximately 20 seconds. The dominant cost is driven by filtering over the variant table, which is largely independent of sample count. In contrast, \textit{cutevariant}'s cold performance scaled markedly (9.7~s at 1\% to 168.9~s at 100\%), a trend that appears as a clear upward slope in the log view and a steep climb in the linear view. \textit{VCFdbR‑file} remained entirely flat (30.1--30.9~s cold), while \textit{vcf‑miner} increased roughly 4.5‑fold. These contrasting trajectories reinforce that annotation‑driven queries are variant‑centric by nature, and that the degree of scaling depends on how each engine resolves joins and applies filters across the variant catalogue  

\paragraph{Genotype retrieval and variant‑level analytical queries (Figure~\ref{fig:placeholder4}).}

For genotype retrieval and variant-level analytical queries (Q4--Q5; Figure~\ref{fig:placeholder4}), the scaling profiles differed between the two workloads. 

Q4 showed clear growth for \textit{vcf2db}, increasing from less than 0.3~s at 1\% to approximately 8.5~s at 100\% under cold cache, with a similar trajectory under warm cache from approximately 0.17~s to 7.9~s. On the logarithmic scale, this appears as a regular upward trend, consistent with increasing sample-dependent genotype retrieval. On the linear scale, however, the absolute runtime remained within the low-second range even at the full dataset size. \textit{cutevariant} achieved lower absolute runtimes for Q4, increasing from approximately 0.08~s to 2.1~s under cold cache and from approximately 0.03~s to 1.9~s under warm cache, while \textit{vcf2db} remained within the same order of magnitude at the largest fraction. \textit{vcf-miner} also increased with dataset fraction, from approximately 1.9~s to 11.4~s under cold cache and from approximately 1.0~s to 3.1~s under warm cache. 

Q5 showed a different profile: despite being a genotype-level query, \textit{vcf2db} remained almost flat across fractions, with cold-cache runtimes close to 45~s and warm-cache runtimes close to 44~s. This flat trajectory is evident in the logarithmic panel and corresponds in the linear panel to a stable absolute runtime cost across the full range of dataset fractions. Similar near-flat behaviour was observed for \textit{cutevariant} in cold cache, although at approximately 70--75~s, while its warm-cache runtime remained much lower, around 6.4--6.6~s. \textit{VCFdbR-file} also remained stable for Q5, with cold-cache values around 25--28~s and warm-cache values close to 14~s, whereas \textit{vcf-miner} increased from approximately 174~s to more than 717~s under cold cache. These results indicate that Q4 followed a more sample-dependent scaling profile, whereas Q5 was dominated by a more stable variant-centric cost for \textit{vcf2db} across dataset fractions.

\paragraph{Aggregation‑heavy workloads (Figure~\ref{fig:placeholder5}).}

Queries Q6 and Q7 represent the most expensive analytical tasks, and their scaling profiles underscore the fixed cost of full‑dataset aggregation. For \textit{vcf2db}, Q6 cold runtimes were 310--316~s at every fraction, with warm times decreasing moderately to 211--248~s. The log‑scale curves are nearly horizontal, and the linear‑scale view shows that runtimes remain firmly in the hundreds‑of‑seconds range regardless of sample count. No meaningful reduction was achieved by reducing the sample set to 1\%, indicating that the cost is dominated by cross‑table joins and multi‑level grouping that must process the entire variant and annotation space, which does not shrink proportionally with sample count. \textit{cutevariant}, while faster than \textit{vcf2db} in absolute terms at low fractions (63.5~s cold at 1\%), showed a pronounced increase with sample number, rising to 740.9~s at 100\%, a trend clearly visible as an upward slope in the log view and a steep climb in the linear view. Its warm runtimes remained constant at approximately 50~s. \textit{VCFdbR‑table} and \textit{VCFdbR‑file} displayed only mild scaling (66.6--136.9~s and 66.2--66.3~s cold, respectively), confirming that their aggregation pathways are largely insensitive to cohort size. \textit{vcf‑miner} does not allow for aggregation, therefore execution times for query Q6 are not reported for this solution.

For Q7 (chromosome‑level summary statistics), \textit{vcf2db} was again remarkably flat, with cold times of 122.7--138.0~s across fractions, and warm times tightly clustered at 119.8--123.1~s. Similarly, \textit{cutevariant} cold (70.9--74.2~s) and \textit{VCFdbR‑file} cold (61.1--62.1~s) exhibited negligible scaling. However, \textit{vcf‑miner} demonstrated clear sample‑dependent growth (171.1~s at 1\% to 726.4~s at 100\% cold). The flat profiles of \textit{vcf2db} and others illustrate that once a query requires a full‑table scan with grouping and aggregate functions over essentially the entire variant catalogue, the addition of more samples does not substantially alter the total data volume being aggregated; the cost is already maximal at the smallest cohort. The dual‑scale representation confirms this: the log view shows slope $\approx$ 0, while the linear view reveals that the absolute penalty remains high even at 1\% scale, effectively precluding interactive use.

Collectively, the scaling analysis reinforces the central observation that query structure is the primary determinant of performance. Variant‑centric retrieval and filtering tasks scale negligibly with cohort size, sample‑driven genotype extraction scales near‑linearly, and aggregation‑centric workloads incur a high fixed cost that is largely independent of the number of samples. For \textit{vcf2db}, this means that its strengths in selective, index‑driven queries are preserved even as cohorts grow, whereas its limitations on analytical aggregation are not simply a matter of dataset scale but are inherent to the computational complexity of the queries themselves.
\section{Discussion}
\label{sec:disc}

The first aim of this work was to provide a modular sample-centric relational model for multi-omics integration and data retrieval. Rather than proposing \textit{vcf2db} as a tool optimized for a single access pattern, we designed it as a generalizable data-management framework in which genomic variation data represent the first implemented and most demanding use case. The results should therefore be interpreted in relation to the three research questions that guided the study: the feasibility of decomposing annotated VCF data into normalized relational entities, the ability of the schema to support modular extension beyond genomics, and the competitiveness of the approach compared with existing VCF-oriented systems.

\subsection{RQ1: Can annotated VCF data be represented relationally while preserving meaningful and efficient genomic retrieval?}

The first research question asked whether annotated VCF data can be decomposed into normalized relational entities in a way that preserves biologically meaningful querying while maintaining efficient retrieval for common genomic tasks. The results support this objective, particularly for selective and annotation-driven workloads. In \textit{vcf2db}, genomic loci, alternative alleles, functional annotations, sample-specific genotypes, and samples are represented as distinct but connected entities. This organization reflects the biological levels at which genomic data are commonly queried: genomic coordinates, allelic variation, functional consequence, and sample-level genotype observations.

The strongest evidence for this design comes from Q1--Q3. These queries targeted genomic intervals, gene-consequence combinations, and rare high-impact variants. Across these workloads, \textit{vcf2db} achieved the lowest or near-lowest execution times among the evaluated systems. This indicates that normalization did not prevent efficient retrieval. On the contrary, the combination of relational decomposition, chromosome-based partitioning, and targeted indexing allowed the database to answer biologically meaningful queries efficiently. The performance observed for Q2 is particularly relevant, because gene- and consequence-based filtering depends directly on how functional annotations are represented and linked to variants. The result suggests that the normalized annotation model supports efficient retrieval over attributes that are central to variant prioritization.

These findings answer RQ1 positively for targeted genomic retrieval. The schema supports common biological access patterns without requiring users to reconstruct relationships externally from flat files. Coordinate-based searches can operate on variant loci, annotation-driven searches can target gene and consequence fields, and genotype-level queries can connect alleles to samples through explicit relational links. This provides a structured and interpretable alternative to file-centric querying, where biological relationships are often implicit and handled by external scripts.

However, the results also show that the answer to RQ1 is workload-dependent. While \textit{vcf2db} performed well for selective retrieval, aggregation-heavy workloads exposed the main limitation of the current implementation. Q6 and Q7 required global summarization, grouping by genes or chromosomes, and large joins across variants and annotations. These queries produced substantially higher runtimes than selective retrieval tasks. The scaling analysis further showed that these costs were largely independent of the number of samples, indicating that the bottleneck lies mainly in scanning and aggregating the variant and annotation space rather than in cohort size alone. Therefore, the current schema is effective for biologically targeted retrieval, but full-dataset analytical aggregation will require additional optimization, such as materialized views, precomputed summaries, denormalized analytical tables, or workload-specific indexes.

\subsection{RQ2: Can the schema be extended modularly beyond genomics?}

The second research question asked whether a sample-centric relational schema can support direct and modular extension beyond genomics without redesigning the original genomic component. The integration of the transcriptomic layer provides a direct answer to this question. Gene-expression data were added through additional tables linked to the same sample identifiers used by the genomic schema. This extension did not require changes to the existing variant, allele, annotation, or genotype tables.

This result confirms the value of placing samples at the centre of the model. In multi-omics studies, different molecular measurements are often generated from the same biological units, but they differ substantially in format, structure, and analytical meaning. A sample-centric design allows each modality to retain its own internal organization while remaining connected to the rest of the database through explicit relational constraints. In this way, the schema separates two requirements that are often conflated: modality-specific representation and cross-modality integration.

The transcriptomic proof of concept therefore answers RQ2 positively. Although the expression data used in this study were synthetic and not intended for biological discovery, they demonstrate that the architecture can evolve incrementally. New omics layers can be added as independent modules, provided that they are linked to the shared sample backbone. In practical terms, this makes it possible to express cross-layer queries directly at the database level. For example, one can retrieve samples carrying variants in a gene of interest and inspect expression values for the same gene or related genes in those samples. This type of query would be more difficult to maintain in file-centric workflows, where integration usually depends on repeated preprocessing, file matching, and modality-specific scripts.

\subsection{RQ3: Does the model remain competitive while providing a stronger basis for cross-layer querying?}

The third research question asked whether the proposed model can remain competitive with existing VCF-oriented systems across representative retrieval workloads while providing a stronger basis for cross-layer querying. The benchmark results suggest a balanced answer. \textit{vcf2db} was highly competitive for selective and annotation-driven queries, remained practical for genotype retrieval, but was less efficient for aggregation-heavy workloads.

For Q1--Q3, \textit{vcf2db} achieved strong performance, often outperforming or closely matching the alternative systems. These results are important because selective genomic retrieval is a common starting point for downstream interpretation and variant prioritization. They show that the relational and sample-centric design does not impose a prohibitive performance penalty for common VCF-derived queries. Instead, it can support efficient access when queries are biologically targeted and can exploit indexes and partition pruning.

For Q4 and Q5, the picture was more nuanced. Q4 returned millions of per-sample genotype records, and its runtime increased as the number of samples grew. This behaviour is expected because genotype retrieval is intrinsically sample-dependent: larger cohorts produce more genotype observations to access and return. Although \textit{vcf2db} was not always the fastest system for this query class, it remained within a practical runtime range while preserving a transparent relational representation of genotype data. This transparency is important for integration, because genotype records can be directly joined with sample metadata and future omics layers.

Aggregation-heavy workloads, represented by Q6 and Q7, were the least favourable case for \textit{vcf2db}. These queries require broad scans and multi-level summarization, which are not the primary target of the current normalized schema. The cache analysis supports this interpretation. Selective retrieval tasks benefited from warm-cache conditions, indicating a relevant I/O component, whereas genotype-level and aggregation-heavy tasks showed limited improvement after cache warming. This suggests that their runtime is dominated by computation, join execution, grouping, and result materialization rather than disk access alone.

Taken together, these results show that \textit{vcf2db} offers a different balance of priorities compared with existing VCF-oriented systems. Tools such as \textit{cutevariant}, \textit{VCFdbR}, and \textit{vcf-miner} may achieve better runtimes for specific workloads, especially when their internal representation is optimized for that task. The contribution of \textit{vcf2db} is not to replace these systems in every scenario, but to combine competitive genomic retrieval with a schema explicitly designed for modular multi-omics integration. In this sense, performance is a necessary condition for usability, while extensibility is the main methodological contribution.

On the other hand, it is important to acknowledge some limitations that define the scope of the present proof of concept. First, the benchmark was performed on the European subset of the 1000 Genomes Project rather than on the complete cohort, because benchmarking the full dataset would require substantially higher computational resources. Although the selected dataset is large enough to provide a demanding and realistic test case for the proposed architecture, further evaluation on larger and more diverse cohorts will be necessary to assess performance at biobank scale. Second, due to the limited availability of public matched multi-omics datasets suitable for this type of controlled evaluation, the transcriptomic layer was generated synthetically. This was appropriate for testing schema extensibility under controlled conditions, but future work should validate the approach using real matched genomic and transcriptomic data, and eventually additional omics layers. Third, the evaluation focused mainly on query runtime because the primary objective of this proof of concept was to determine whether a modular and extensible relational architecture could remain competitive with existing VCF-oriented solutions for representative retrieval workloads. Establishing comparable retrieval performance was a necessary first step before assessing broader operational dimensions of the system, since the main contribution of \textit{vcf2db} lies in combining practical genomic querying with schema-level expandability toward multi-omics integration. Other metrics, including ingestion time, storage footprint, concurrent access, update operations, user-facing query interfaces, and integration with downstream analysis pipelines, remain highly relevant for real-world deployment. However, these aspects depend not only on the schema design, but also on implementation choices, database configuration, hardware environment, and intended usage scenario. For this reason, they should be evaluated in future work as part of a more comprehensive system-level assessment.

Future development should therefore proceed in two directions. 
For aggregation-intensive analyses, the normalized relational schema could be complemented by derived analytical structures, such as materialized views, precomputed variant-, gene-, chromosome-, and sample-group-level statistics, and workload-specific indexing strategies for frequently filtered, joined, and aggregated attributes. Where appropriate, hybrid architectures incorporating columnar or other analytical storage technologies could also be investigated, while retaining the relational schema as the authoritative layer for data integration and consistency.
This would allow the system to be tested not only for structural extensibility, but also for biologically meaningful cross-layer analyses.

Overall, the results answer the three research questions in a coherent way. Annotated VCF data can be decomposed into normalized relational entities while preserving efficient and biologically meaningful retrieval, especially for selective coordinate- and annotation-driven queries. A sample-centric schema can be extended modularly beyond genomics, as demonstrated by the transcriptomic proof of concept. Finally, \textit{vcf2db} remains competitive for important genomic retrieval workloads while providing a stronger basis for cross-layer querying, although aggregation-heavy analysis remains the main target for future optimization. These findings support \textit{vcf2db} as a modular relational methodology for scalable and extensible omics data management.

\section{Conclusion}
\label{sec:conc}

Modern omics research increasingly depends on the ability to connect large, heterogeneous, and continuously expanding molecular datasets. In this context, the central challenge is not only to store large files, but to organize biological information so that it remains searchable, interpretable, reusable, and extensible across multiple layers of evidence. File-centric workflows, although still widely used, become progressively less suitable when analyses require the systematic integration of genomic, transcriptomic, phenotypic, and other omics measurements across shared biological samples.

This work was motivated by the need for a structured way to manage and query large-scale VCF data. It also addressed key
bioinformatics questions related to relational VCF modelling, modular multi-omics integration, and performance compared with
existing VCF-oriented systems. To meet these aims, this study introduced vcf2db, a relational and sample-centric methodology for
organizing, managing, and querying large-scale omics data.
Genomic variation data were used as the initial proof of concept because annotated VCF files represent a demanding test case in terms of size, structure, annotation complexity, and sample-level dimensionality. By decomposing VCF information into normalized relational entities, including variants, alternative alleles, annotations, genotypes, and samples, \textit{vcf2db} provides a structured representation of genomic variation while preserving explicit links between molecular observations and biological samples.

The experimental results show that this design is both operationally practical and methodologically extensible. Across representative genomic retrieval workloads, \textit{vcf2db} achieved strong performance for selective coordinate-based and annotation-driven queries, demonstrating that relational normalization can support efficient access to biologically meaningful variant subsets. Genotype-level retrieval remained practical, while aggregation-heavy workloads highlighted the main area for future optimization. Together, these findings show that the proposed model can manage large annotated genomic datasets without sacrificing the flexibility required for integrative analysis.

Beyond retrieval performance, the main contribution of \textit{vcf2db} lies in its treatment of multi-omics integration as a database-design principle. The extension of the schema with transcriptomic data demonstrated that new molecular layers can be incorporated without redesigning the existing genomic component. By using samples as the central integration backbone, each omics modality can retain its own internal structure while remaining directly linkable to other layers through shared sample identifiers. This enables cross-layer retrieval tasks to be expressed directly as database queries, reducing dependence on fragmented files, repeated preprocessing, and modality-specific scripts.

At the same time, the scope of this proof of concept should be considered when interpreting the results. The benchmark was performed on the European subset of the 1000 Genomes Project rather than on the complete cohort, the transcriptomic layer was synthetically generated due to the limited availability of suitable public matched multi-omics datasets, and the evaluation focused primarily on query runtime. These choices were appropriate for assessing the core feasibility of the proposed architecture, but future work should extend the evaluation to larger and more diverse cohorts, real matched multi-omics datasets, and additional operational metrics such as ingestion time and storage footprint.

Overall, \textit{vcf2db} bridges two requirements that are often addressed separately: efficient access to large-scale genomic variation data and modular organization for multi-omics integration. Future developments will focus on improving aggregation-heavy analytical workloads, evaluating the framework on larger and real matched multi-omics cohorts, and extending the schema toward additional molecular and clinical data layers. By combining relational modelling, sample-centric integration, competitive retrieval performance, and modular expandability, \textit{vcf2db} moves omics data management from passive storage toward an integrated infrastructure where biological complexity can be queried, connected, and reused.

\backmatter

\section*{Declarations}

\subsection*{Ethics approval and consent to participate}
Not applicable. This study did not involve human participants or animals, and all data used were publicly available or synthetically generated.

\subsection*{Consent for publication}
Not applicable.

\subsection*{Availability of data and materials}
The genomic data used in this study are publicly available from the 1000 Genomes Project Phase 3 release. The raw data results are available at \url{https://github.com/BioGeMT/variant_db_benchmarking/tree/main/data/queries}. The code for data ingestion is available at \url{https://github.com/BioGeMT/vcf2db}.

\subsection*{Competing interests}
The authors declare that they have no competing interests.

\subsection*{Funding}
This work has been supported by funding from the project BioGeMT (HORIZON-WIDERA-2022 Grant ID: 101086768) at the University of Malta.

\subsection*{Authors' contributions}
AB and PA conceived the study. AB and DF designed and developed the methodology. DF and AG contributed to the software and maintenance of the repository. AB and AG contributed to validation and findings. AB, DG and AG wrote the first draft of the manuscript. AB finalized the manuscript. All authors read and approved the final manuscript.

\subsection*{Acknowledgements}
Not applicable.

\subsection*{Authors' information}
Not applicable.

\bibliography{./sections/bibliography/bibliography.bib}% common bib file

\end{document}